

\input phyzzx

\nopubblock
\line{\hfil JHU-TIPAC-930002}
\line{\hfil January, 1993}
\titlepage
\title{{\seventeenrm
Third Generation Effects on Fermion Mass Predictions in
Supersymmetric Grand Unified Theories}\footnote\dag{
This work has been supported by the NSF under grant PHY-90-96198.}}
\author{   Stephen G. Naculich\footnote\ast{
        NACULICH @ CASA.PHA.JHU.EDU }}
\address{Department of Physics and Astronomy     \break
        The Johns Hopkins University             \break
        Baltimore, MD 21218}
\abstract{
Relations among fermion masses and mixing angles at the scale
of grand unification are modified at lower energies by
renormalization group running induced by gauge and Yukawa
couplings.   In supersymmetric theories, the $b$ quark and
$\tau$ lepton Yukawa couplings, as well as the $t$ quark
coupling, may cause significant running if $\tan \beta$,
the ratio of Higgs field expectation values, is large.
We present approximate analytic expressions for the
scaling factors for fermion masses and CKM matrix elements
induced by all three third generation Yukawa couplings.
We then determine how running caused by the third generation
of fermions affects the predictions arising from three
possible forms for the Yukawa coupling matrices at the GUT
scale: the Georgi-Jarlskog, Giudice, and Fritzsch textures.
}
\endpage

\overfullrule=0pt

%
\def\PLB{ \sl Phys. Lett.         \bf B}

\def\NPB{ \sl Nucl. Phys.         \bf B}

\def\PRD{ \sl Phys. Rev.          \bf D}
\def\PRL{ \sl Phys. Rev. Lett.    \bf  }
\def\PRp{ \sl Phys. Rep.          \bf  }
\def\ZPC{ \sl Zeit. Phys.         \bf C}

\def\MPLA{ \sl Mod. Phys. Lett. \bf  A}
\def\TNYAS{ \sl Trans. N. Y. Acad. Sci. \bf}
\def\ARNPS{ \sl Annu. Rev. Nucl. Part. Sci. \bf}
\def\PTP{ \sl Prog. Theor. Phys. \bf }
\def\viz{{\it viz.}}

\def\third{{1\over 3}}
\def\twothirds{{2\over 3}}
\def\d { {\rm d} }
\def\e { {\rm e} }

\def\gsim{\mathrel{\raise.3ex\hbox{$>$\kern-.75em\lower1ex\hbox{$\sim$}}}}
\def\lsim{\mathrel{\raise.3ex\hbox{$<$\kern-.75em\lower1ex\hbox{$\sim$}}}}
\def\ts {\textstyle}
\def\dag{ \dagger }

\def\ddt{ {\d \over \d t} }
\def\eps{ \varepsilon }
\def\scale{\mu}
\def\pscale { \scale^\prime }
\def\bscale{ {\overline\scale} }
\def\msusy {\scale_{\rm susy} }
\def\bg { \overline{g} }

\def\btheta{ \overline{\theta} }
\def\MF { Y_F }
\def\MU { Y_U }
\def\MD { Y_D }
\def\ME { Y_E }
\def\Yhat{ \hat{Y} }
\def\MFd { \Yhat_F }
\def\MUd { \Yhat_U }
\def\MDd { \Yhat_D }
\def\MEd { \Yhat_E }
\def\FL { F_L }

\def\bUL{ \overline{U}_L }
\def\bDL{ \overline{D}_L }
\def\bEL{ \overline{E}_L }
\def\FR { F_R }
\def\UR { U_R }
\def\DR { D_R }
\def\ER { E_R }
\def\LF { {L_F} }
\def\LU { {L_U} }
\def\LD { {L_D} }
\def\LE { {L_E} }
\def\LFdot { \dot{L}_F }
\def\LUdot { \dot{L}_U }
\def\LDdot { \dot{L}_D }
\def\LEdot { \dot{L}_E }
\def\RF { {R_F} }
\def\RU { {R_U} }
\def\RD { {R_D} }

\def\Vdag { {V^\dag} }
\def\s { {\rm s} }
\def\c { {\rm c} }
\def\bs { \overline{\rm s} }
\def\bc { \overline{\rm c} }
\def\sone{ \s_1 }
\def\stwo{ \s_2 }
\def\sthr{ \s_3 }

\def\cone{ \c_1}
\def\ctwo{ \c_2}
\def\cthr{ \c_3 }

\def\bsone{ \bs_1 }
\def\bstwo{ \bs_2 }
\def\bsthr{ \bs_3 }
\def\bsthrp{ \bs^\prime_3 }
\def\bsthrpp{ \bs^{\prime\prime}_3 }
\def\bcone{ \bc_1}
\def\bctwo{ \bc_2}
\def\bcthr{ \bc_3 }
\def\bcthrp{ \bc^\prime_3 }
\def\bcthrpp{ \bc^{\prime\prime}_3 }
\def\phase{ \e^{i\phi} }

\def\Vij{ V_{\alpha\beta} }
\def\Vcb{ \left| V_{cb} \right| }
\def\Vub{ \left| V_{ub} \right| }
\def\Vus{ \left| V_{us} \right| }
\def\bV{ \overline{V} }
\def\bVcb{ \left| \overline{V}_{cb} \right| }
\def\bVub{ \left| \overline{V}_{ub} \right| }
\def\bVus{ \left| \overline{V}_{us} \right| }
\def\bVij{ \overline{V}_{\alpha\beta} }
\def\lam { y }
\def\blam{ \overline{\lam} }
\def\Bt { B_t}
\def\Bb { B_b}
\def\Btau { B_{\tau} }
\def\tb{ \tan\beta}
\def\GeV{ {\rm~GeV} }
\def\MeV{ {\rm~MeV} }
\def\RG{RG~}

\chapter{Introduction}

Thirteen is the number of independent parameters
required to describe the fermion sector of the standard model:
nine masses, three mixing angles, and a CP-violating phase.
Although such a plethora of arbitrary parameters is
usually regarded as a weakness,
we could instead view the situation as an opportunity
to reach beyond the standard model.
Because the fermion parameters can take on
arbitrary values in the standard model,
any prediction of these parameters can only
come from beyond the standard model.
Conversely, their observed experimental values
could provide a clue to new physics.

\REF\rrSUfive{
H.~Georgi and S.~Glashow,
\PRL 32 \rm (1974) 438; \nextline
M.~S.~Chanowitz, J.~Ellis, and M.~K.~Gaillard,
\NPB 128 \rm (1977) 506; \nextline
A.~Buras, J.~Ellis, M.~K.~Gaillard, and D.~Nanopoulos,
\NPB 135 \rm (1978) 66.
}
\REF\rrGJ{
H.~Georgi and C.~Jarlskog,
\PLB 86 \rm (1979) 297.
}
\REF\rrHRR{
J.~Harvey, P.~Ramond, and D.~Reiss,
\PLB  92 \rm (1980) 309;
\NPB  199 \rm (1982) 223.
}
\REF\rrDisc{
S.~Weinberg,
\TNYAS 38 \rm (1977) 185; \nextline
F.~Wilczek and A.~Zee,
\PLB  70 \rm (1977) 418.
}
\REF\rrFri{
H.~Fritzsch,
\PLB 70 \rm (1977) 436;
\bf 73 \rm (1978) 317;
\bf 166 \rm (1986) 423.
}
A step in this direction has been taken with the
discovery of various phenomenological relations
among fermion masses and mixing angles,
which could be viewed as modern-day Balmer formulae.
Two types of relations among fermion parameters
have been explored.
The first type links the masses of fermions
within the same generation to one another.
Such relations result naturally
from grand unified theories (GUTs)
when the fermions belong
to a common grand unified representation
and couple to a single Higgs field;
group theory then dictates a relation between their masses
[\rrSUfive--\rrHRR].
The second type of relation
connects Cabibbo-Kobayashi-Maskawa (CKM) matrix elements
with ratios of masses of fermions in different generations.
These relations arise naturally when
certain entries of the Yukawa matrices vanish,
perhaps as a result of discrete symmetries [\rrDisc, \rrFri].

These various relations among Yukawa couplings
are presumably a consequence of new physics,
and therefore hold at the energy scale of the new physics,
\eg, the grand unification scale.
But Yukawa couplings evolve
in accord with the renormalization group (RG) equations,
so relations among them that apply at one scale
will not necessarily hold at another scale.
Therefore, before they can be compared with low-energy data,
GUT-scale relations must be corrected to account for \RG running.

\REF\rrCEL{
T.~P.~Cheng, E.~Eichten, and L.-F.~Li,
\PRD 9 \rm (1974) 2259.
}
\REF\rrIKKT{
K.~Inoue, A.~Kakuto, H.~Komatsu, and S.~Takeshita,
\PTP 67 \rm (1982) 1889.
}
\REF\rrIL{
L.~Iba\~nez and C.~Lopez,
\NPB  233 \rm (1984) 511.
}
\REF\rrMP{
E.~Ma and S.~Pakvasa,
\PLB  86 \rm (1979) 43;
\PRD 20 \rm (1979) 2899; \nextline
K.~Sasaki,
\ZPC  32 \rm (1986) 149; \nextline
K.~S.~Babu,
\ZPC  35 \rm (1987) 69.
}
\REF\rrFla{
H.~Arason, D.~Casta\~no, B.~Keszthelyi, S.~Mikaelian, E.~Piard,
P.~Ramond, and B.~Wright,
\PRL 67 \rm (1991) 2933; \nextline
P.~Ramond,
Florida preprint UFIFT-92-4; \nextline
H.~Arason, D.~Casta\~no, E.~Piard, and P.~Ramond,
Florida preprint UFIFT-92-8.
}
\REF\rrKLN{
S.~Kelley, J.~Lopez, and D.~Nanopoulos,
\PLB 274 \rm (1992) 387.
}
\REF\rrDHR{
S.~Dimopoulos, L.~J.~Hall, and S.~Raby,
\PRL  68 \rm (1992) 1984;
\PRD  45 \rm (1992) 4192.
}
\REF\rrBBHZ{
V.~Barger, M.~S.~Berger, T.~Han, and M.~Zralek,
\PRL 68 \rm (1992) 3394.
}
\REF\rrGiu{
G.~Giudice,
\MPLA 7 \rm (1992) 2429.
}
Most of the running of the Yukawa couplings
is induced by the gauge couplings.
For example, group-theoretic relations between
quark and lepton masses at the GUT scale
are greatly modified at low-energy scales
because quarks and leptons have different gauge couplings
and their masses therefore run differently [\rrSUfive--\rrHRR].
If they are sufficiently large, however,
Yukawa couplings themselves induce further running [\rrCEL],
and therefore further modifications of relations
between  masses [\rrIKKT, \rrIL].
Yukawa couplings also induce running
of the CKM matrix elements [\rrMP],
which are invariant under gauge coupling-induced running.
The effect of the top quark Yukawa coupling
on fermion mass and mixing angle relations in supersymmetric theories
was recently investigated in refs.~[\rrFla--\rrGiu].

\REF\rrBBO{
V.~Barger, M.~S.~Berger, and P.~Ohmann,
Madison preprint MAD/PH/711.
}
\REF\rrADHR{
G.~Anderson, S.~Dimopoulos, L.~J.~Hall, and S.~Raby,
preprint OHSTPY-HEP-92-018.
}
\REF\rrBS{
K.~S.~Babu and Q.~Shafi,
Bartol preprint BA-92-70.
}
In the standard model,
the Yukawa couplings of fermions other than the top quark
are too small to cause significant running.
In supersymmetric theories, however,
the Yukawa couplings of the other third generation fermions,
the $b$ quark and $\tau$ lepton,
may be comparable to the $t$ quark Yukawa coupling,
even though their masses are much smaller.
This occurs when the expectation value of the Higgs field
to which $b$ and $\tau$ are coupled
is much less than that of the Higgs field
to which $t$ is coupled;
that is, when $\tb$, the ratio of expectation values
of the two Higgs fields, is large.
In this regime, all three third generation
fermions may cause significant running.
This case has also been investigated recently
by several authors [\rrFla, \rrKLN, \rrBBO--\rrBS],
who solved the \RG equations numerically.

In this paper, we would like to calculate the effect
of \RG running induced by the entire third generation of fermions
{\it analytically}.
An analytic result would have several advantages over a numerical solution.
In addition to enhancing intuition
about the effects of Yukawa coupling-induced running,
it would allow one to see transparently
how changes in the input parameters affect the predictions,
without having to re-run the numerical routines
for each new set of data.
It would also simplify the error analysis.
Although it is not possible to solve the \RG equations exactly,
we introduce an approximation (good to within a few percent)
that includes the running
induced by all three third generation fermions
and that allows an analytic solution.
We then use this approximate solution
to determine the effects of
\RG running on several different sets of mass and mixing angle relations,
and on the predictions that follow from those relations.

The \RG running of the
Yukawa couplings
is logically independent
of the fermion mass and mixing angle relations
because the latter arise from new physics at the GUT scale
whereas the running from the GUT scale to the low-energy scale
only depends on the particle spectrum below the GUT scale.
{}From the point of view of the \RG equations,
the only role of the fermion relations
is to provide boundary conditions at the GUT scale.
Because of this,
we will be able to analyze the \RG running of the Yukawa couplings
independently of any particular set of fermion relations.

One approach is to evolve the matrices of Yukawa couplings
down to the low-energy scale,
and then diagonalize them to find the fermion masses and
mixing angles.
Many degrees of freedom of the Yukawa matrices are not physical,
however, because of the freedom
to perform unitary redefinitions of the fermion bases.
Therefore, we instead begin with the \RG equations for the
smaller set of physical parameters,
\viz, the fermion masses, mixing angles, and CP-violating phase.
This simplifies the computational task
by reducing the number of equations,
and allows us to deal only with
physical quantities throughout.

In sect.~2, we review the one-loop \RG equations
for fermion masses and CKM matrix elements
in the minimal supersymmetric standard model.
Adopting a non-standard parametrization of the CKM matrix,
we obtain explicit \RG equations for the mixing angles
and CP-violating phase.
In sect.~3, we introduce an approximation
that allows us to solve the \RG equations analytically,
including the effects of the entire third generation of fermions.
We then analyze in sect.~4
the \RG effects on
three different sets of fermion mass and mixing angle relations
that might result from new physics at the GUT scale.
Sect.~5 contains our conclusions.

\chapter{Running Masses and Mixing Angles}

In this section, we review
the renormalization group running
of fermion masses and CKM matrix elements.
We also include
a discussion of phase choices for the CKM matrix
so that we can obtain explicit \RG equations for
the mixing angles which parametrize it.

\REF\rrSUSY{
J.~Ellis, S.~Kelley, and D.~Nanopoulos,
\PLB 249 \rm (1990) 441; \nextline
U.~Amaldi, W.~de Boer, and H.~F\"urstenau,
\PLB 260 \rm (1991) 447; \nextline
P.~Langacker and M.~Luo,
\PRD 44 \rm (1991) 817.
}
\REF\rrPR{
B.~Pendleton and G.~G.~Ross,
\PLB 98 \rm (1981) 291.
}
As noted in the introduction,
the renormalization group analysis
is independent of the new physics
responsible for relations
between fermion masses and mixing angles,
and can therefore be applied to different sets of relations.
We assume, however,
that the relations result from some grand unified theory,
and therefore that
the three gauge couplings meet at a single scale $\bscale$.
This can be achieved
in the context of
the minimal supersymmetric standard model,
with the supersymmetry breaking scale $\msusy$
between 100 GeV and 10 TeV [\rrSUSY].
We will assume that this framework describes physics up to
the GUT scale $\bscale$.

The one-loop \RG equations
for the gauge couplings
are
$$
16\pi^2 \ddt \ln g_i = b_i g_i^2 ,
\qquad\qquad
t = \ln  \scale,
\eqn\eeGaugeRGE
$$
and have solutions
$$
{1 \over g^2_i (\scale)} =
{1 \over \bg^2}  - {b_i\over 8 \pi^2} \ln \left( \scale \over \bscale \right).
\eqn\eeGaugeSoln
$$
Between $\msusy$ and the grand unified scale $\bscale$,
the coefficients are given by
$$
(b_1, b_2, b_3) = \left( {\ts  {33\over 5}, 1, -3  } \right).
\eqn\eeGaugeRGECoeff
$$
We choose $\msusy = 170$ GeV for convenience
(close to the top quark mass);
our results will be rather insensitive to the exact
value of $\msusy$.
Using [\rrBBO]
$$
{g_1^2 (\msusy) \over 4\pi} = {1\over 58.5}, \qquad\qquad
{g_2^2 (\msusy) \over 4\pi} = {1\over 30.1}, \qquad\qquad
{\rm for~} \msusy = 170 \GeV,
\eqn\eeGaugeBC
$$
we obtain
$$
{\bg^2 \over 4\pi } = {1\over 25.0},
\qquad\qquad
\bscale = 1.2 \times 10^{16} \GeV.
\eqn\eeGaugeGUT
$$
Throughout this paper, an overline denotes
quantities evaluated at the GUT scale.

In the minimal supersymmetric standard model,
the charge $\twothirds$ quarks
couple to a Higgs field with
expectation value
$ (v / \sqrt 2)  \sin \beta$,
where $ v = 246$ GeV and $\beta$ is arbitrary;
the charged leptons and charge $-\third$ quarks
couple to a Higgs field with
expectation value
$ (v / \sqrt 2)  \cos \beta$.
The fermion masses come from the Yukawa couplings
$$
L_{\rm Yuk}
    = \left( {v\over\sqrt 2} \sin \beta\right) \bUL \MU \UR
    + \left( {v\over\sqrt 2} \cos \beta\right) \bDL \MD \DR
    + \left( {v\over\sqrt 2} \cos \beta\right) \bEL \ME \ER
    + {\it h.c.} ,
\eqn\eeYukLag
$$
where
$$
       U =  \pmatrix{  u \cr
                  c \cr
                  t \cr },
\qquad \qquad D =  \pmatrix{d \cr
                  s \cr
                  b \cr },
\qquad \qquad E =  \pmatrix{e \cr
                  \mu \cr
                  \tau \cr },
\eqn\eeFermVec
$$
are the fermion fields,
and
$\MU$, $\MD$, and $\ME$
are arbitrary complex $ 3 \times 3 $ matrices.
(In this paper, we take the neutrinos to be massless.)
These Yukawa matrices obey
the one-loop supersymmetric \RG equations [\rrIKKT]
$$
\eqalign{
  16 \pi^2 \ddt \MU
&
  ~=~\left[ -~c^u_i g_i^2 + \Tr ( 3 \MU {\MU}^\dag )
  + 3 \MU {\MU}^\dag + \MD {\MD}^\dag \right] \MU ,
\cr
  16 \pi^2 \ddt \MD
&
   ~=~\left[ -~c^d_i g_i^2 + \Tr ( 3 \MD {\MD}^\dag + \ME {\ME}^\dag )
   + 3 \MD {\MD}^\dag + \MU {\MU}^\dag \right] \MD ,
\cr
   16 \pi^2 \ddt \ME
&
   ~=~\left[-~c^e_i g_i^2 + \Tr ( 3 \MD {\MD}^\dag + \ME {\ME}^\dag )
  + 3 \ME {\ME}^\dag \right] \ME  ,
\cr
}
\eqn\eeYukRGE
$$
where
$$
 (c^u_1, c^u_2, c^u_3)
= \left( {\ts {13\over 15}, 3, {16\over 3} } \right),
\qquad
 (c^d_1, c^d_2, c^d_3)
= \left( {\ts {7\over 15}, 3, {16\over 3} } \right),
\qquad
 (c^e_1, c^e_2, c^e_3)
= \left( {\ts {27\over 15}, 3, 0} \right) ,
\eqn\eeYukRGECoeff
$$
between $\msusy$ and $\bscale$.

Not all the parameters of the Yukawa matrices
are physical.
Under an arbitrary unitary transformation
on the fermion bases,
$ \FL \to \LF \FL$,
$ \FR \to \RF \FR$
(where $F = U$, $D$, $E$),
the Yukawa matrix undergoes a bi-unitary transformation,
$ \MF  \to \LF^\dag \MF \RF $,
and the charged current becomes off-diagonal,
with mixing matrix $ \LU^\dag \LD$.
We may perform scale-dependent unitary transformations
$\LF (\scale)$ and $\RF (\scale)$
on the fermion bases
so as to diagonalize the Yukawa matrices at each scale.
Thus
$$
\MFd (\scale) = \LF^\dag (\scale) \MF (\scale) \RF(\scale),
\qquad\qquad
F = U, D, E,
\eqn\eeYukDiag
$$
where $\MFd$ denotes the diagonalized Yukawa matrix, and
$$
V (\scale) = \LU^\dag (\scale) \LD (\scale)
\eqn\eeCKMDef
$$
is the corresponding scale-dependent CKM matrix.

We now derive \RG equations for
the physically relevant quantities:
the Yukawa eigenvalues $\MFd (\scale)$
and the CKM matrix $V(\scale)$ [\rrMP].
The transformations on the right-handed fields
are irrelevant to the CKM matrix,
so we begin by writing $\MFd$ in terms of $\LF$ only
$$
\MFd^2  (\scale)
= \LF^\dag (\scale) \MF (\scale) {\MF}^\dag (\scale) \LF (\scale),
\qquad\qquad
F = U, D, E.
\eqn\eeYukSqDiag
$$
Differentiating eq.~\eeYukSqDiag,
and using eqs.~\eeYukRGE, we obtain
$$
\eqalign{
\ddt \left( \MUd^2 \right)
&
= [ \MUd^2, \LU^\dag \LUdot ]
+ {1\over 16\pi^2} \left[ 6 \Tr (\MUd^2) \MUd^2 + 6 \MUd^4
+ V \MDd^2 V^\dag \MUd^2 + \MUd^2 V \MDd^2 V^\dag \right] ,
\crr
\ddt \left( \MDd^2 \right)
&
= [ \MDd^2, \LD^\dag \LDdot ]
 + {1\over 16\pi^2} \Big[ 2 \Tr ( 3\MDd^2 + \MEd^2 ) \MDd^2 + 6 \MDd^4
+ V^\dag \MUd^2 V \MDd^2 + \MDd^2 V^\dag \MUd^2 V \Big] ,
\crr
\ddt \left( \MEd^2 \right)
&
= [ \MEd^2, \LE^\dag \LEdot ]
+  {1\over 16\pi^2} \left[ 2 \Tr ( 3\MDd^2 + \MEd^2 ) \MEd^2 + 6 \MEd^4
\right],
\cr
}
\eqn\eeYukSqRGE
$$
where
$\LFdot = ( \d \LF / \d t )$.
The commutator
$ [ \MFd^2, \LF^\dag \LFdot ] $
has vanishing diagonal elements
because $\MFd^2$ is diagonal.
Thus the \RG equations
for the Yukawa eigenvalues $\lam_\alpha$
follow immediately from the diagonal entries of eqs.~\eeYukSqRGE.
The remaining entries of eqs.~\eeYukSqRGE~yield equations
for the {\it off-diagonal} elements of
$ \LU^\dag \LUdot  $
and
$ \LD^\dag \LDdot  $,
as long as there are no degeneracies among the quark masses:
$$
\eqalign{
( \LU^\dag \LUdot )_{\alpha\beta}
&= {1\over 16\pi^2} \sum_{\gamma=d,s,b}
 {\lam_\beta^2 + \lam_\alpha^2 \over \lam_\beta^2 - \lam_\alpha^2}
   V_{\alpha\gamma} \; \lam_\gamma^2 \; \Vdag_{\gamma\beta},
\qquad \alpha \neq \beta ,
\qquad \alpha, \beta = u,c,t,  \cr
( \LD^\dag \LDdot )_{\alpha\beta}
&  = {1\over 16\pi^2} \sum_{\gamma=u,c,t}
  {\lam_\beta^2 + \lam_\alpha^2 \over \lam_\beta^2 - \lam_\alpha^2 }
   \Vdag_{\alpha\gamma} \; \lam_\gamma^2 \; V_{\gamma\beta} ,
\qquad \alpha \neq \beta,
\qquad \alpha, \beta = d,s,b. \cr
}
\eqn\eeOffdiagUnitRGE
$$
The {\it diagonal} elements of
$ \LU^\dag \LUdot  $
and
$ \LD^\dag \LDdot  $
are {\it not} determined by eqs.~\eeYukSqRGE.
It is easy to see why.
Equation \eeYukSqDiag~determines
$\LU$ and $\LD$
only up to right multiplication
by a diagonal matrix of (scale-dependent) phases.
These undetermined phases contribute
arbitrary imaginary functions
to the diagonal elements of
$ \LU^\dag \LUdot  $
and
$ \LD^\dag \LDdot  $.
(The off-diagonal elements
are unambiguously determined
because they receive no contribution from the phases.)
We can, however, {\it choose} the phases
to make the diagonal elements of
$ \LU^\dag \LUdot  $ and $ \LD^\dag \LDdot  $,
which are manifestly imaginary,
vanish:
$$
( \LU^\dag \LUdot )_{\alpha\alpha}
= ( \LD^\dag \LDdot )_{\alpha\alpha}
= 0  \qquad {\rm ~by~an~appropriate~choice~of~phases.}
\eqn\eeDiagUnitRGE
$$
The \RG equations
for the CKM matrix elements \eeCKMDef~are then [\rrMP]
$$
\eqalign{
16 \pi^2 \ddt V_{\alpha\beta}
& = 16 \pi^2 \left( V \LD^\dag \LDdot - \LU^\dag \LUdot V \right)_{\alpha\beta}
\cr
&
=
 \sum_{\gamma=u,c,t}
 \sum_{\delta=d,s,b \atop \delta \neq \beta}
  {\lam_\beta^2 + \lam_\delta^2 \over \lam_\beta^2 - \lam_\delta^2 }
   V_{\alpha\delta} \Vdag_{\delta\gamma} \; \lam_\gamma^2 \; V_{\gamma\beta}
+
 \sum_{\gamma=u,c,t \atop \gamma \neq \alpha}
 \sum_{\delta=d,s,b }
 { \lam_\alpha^2 + \lam_\gamma^2 \over \lam_\alpha^2 - \lam_\gamma^2 }
   V_{\alpha\delta} \; \lam_\delta^2 \; \Vdag_{\delta\gamma} V_{\gamma\beta},
\cr
}
\eqn\eeCKMRGEone
$$
as long as we choose the phases to guarantee eqs. \eeDiagUnitRGE.

We now neglect the contributions to the running
caused by the first and second generation Yukawa
couplings.
If we further assume
$\lam_u^2 \ll \lam_c^2 \ll \lam_t^2$
and
$\lam_d^2 \ll \lam_s^2 \ll \lam_b^2$,
then the CKM matrix \RG equations \eeCKMRGEone~reduce to
$$
16 \pi^2 \ddt V_{\alpha\beta}
=     \lam_t^2   \sum_{\delta=d,s,b}
    \eps_{\delta\beta} V_{\alpha\delta} \Vdag_{\delta t} V_{t\beta}
  ~+~\lam_b^2  \sum_{\gamma=u,c,t}
    \eps_{\gamma\alpha} V_{\alpha b} \Vdag_{b \gamma} V_{\gamma \beta} ,
\qquad
\eps_{\alpha\beta} = \cases{1  & if $\lam_\alpha < \lam_\beta$,\cr
       				0  & if $\lam_\alpha = \lam_\beta$,\cr
				-1 & if $\lam_\alpha > \lam_\beta$.\cr}
\eqn\eeCKMRGEtwo
$$
The CKM matrix elements $V_{\alpha\beta}$
are not all independent  because of the constraint of unitarity.

We prefer to go a step further than previous treatments
by deriving \RG equations for
a set of {\it independent} quantities
parametrizing the unitary CKM matrix:
the mixing angles and CP-violating phase
(but see ref.~[\rrPR]).
To do so, however, we must squarely face
the issue of phases multiplying the matrix.
We adopt the following
(nonstandard) parametrization of the CKM matrix
$$
V  (\scale) =
\pmatrix{ \e^{i\phi_u} & 0 & 0 \cr
          0 & \e^{i\phi_c} & 0 \cr
	  0 & 0 & \e^{i\phi_t} \cr   }
\pmatrix{ \sone \stwo \cthr + \cone \ctwo \phase &
	  \cone \stwo \cthr - \sone \ctwo \phase &
          \stwo \sthr \cr
          \sone \ctwo \cthr - \cone \stwo \phase &
	  \cone \ctwo \cthr + \sone \stwo \phase &
	  \ctwo \sthr \cr
          -\sone \sthr &
	  -\cone \sthr &
 	  \cthr \cr}
\pmatrix{ \e^{i\phi_d} & 0 & 0 \cr
          0 & \e^{i\phi_s} & 0 \cr
	  0 & 0 & \e^{i\phi_b} \cr   },
\eqn\eeCKMParam
$$
where $\c_i = \cos \theta_i$ and $\s_i = \sin \theta_i$,
and all the parameters are functions of the scale $\scale$.
The middle matrix is chosen
to have real elements in the third row and column.
We cannot automatically eliminate the left and right
phase matrices by rephasing the quark fields;
we have already used that freedom to ensure eqs.~\eeDiagUnitRGE,
and those equations implicitly determine
the functions $\dot{\phi}_\alpha (\scale)$.
We can, however,
impose the boundary condition $ \phi_\alpha (\bscale) = 0 $;
\ie, the initial values
for the five matrix elements $V_{\alpha\beta}$
in the third row or column may be chosen to be real.
One can show,
using  $ V^\dag V = 1$,
that the \RG equations \eeCKMRGEtwo~for this set of five matrix elements
close on themselves.
Since their initial values are real
and the coefficients in the equations are real,
these five matrix elements remain real
(\ie,  $ \phi_\alpha (\scale) = 0 $)
for all $\scale$.
In other words,
the choice of quark phases that guarantees
eqs.~\eeDiagUnitRGE~also implies that the phases
$\phi_\alpha (\scale)$ in the CKM parametrization
\eeCKMParam~vanish
(in the approximation that eq.~\eeCKMRGEtwo~is valid).
We then obtain the \RG equations
for the angles $\theta_i$ and the CP-violating phase $\phi$
by substituting eq.~\eeCKMParam~into eq.~\eeCKMRGEtwo
$$
\eqalign{
16 \pi^2 \ddt \ln \tan \theta_1  & = - \lam_t^2 \sin^2 \theta_3 ,
\qquad \qquad
16 \pi^2 \ddt \ln \tan \theta_3  = - \lam_t^2 - \lam_b^2 ,
\crr
16 \pi^2 \ddt \ln \tan \theta_2 & = - \lam_b^2 \sin^2 \theta_3 ,
\qquad \qquad
16 \pi^2 \ddt \phi = 0.
\cr
}
\eqn\eeAngleRGE
$$
The \RG equations
for the Yukawa eigenvalues [\rrIKKT, \rrMP],
$$
\eqalign{
16\pi^2 \ddt \ln \lam_u
& = -c^u_i g_i^2 + 3\lam_t^2 + \lam_b^2 \cos^2 \theta_2 \sin^2 \theta_3 ,\cr
16\pi^2 \ddt \ln \lam_c
& = -c^u_i g_i^2 + 3 \lam_t^2 + \lam_b^2 \sin^2 \theta_2 \sin^2 \theta_3 ,\cr
16\pi^2 \ddt \ln \lam_t
& = -c^u_i g_i^2 + 6 \lam_t^2 + \lam_b^2 \cos^2 \theta_3 ,\cr
16\pi^2 \ddt \ln \lam_d
& = -c^d_i g_i^2 + \lam_t^2 \sin^2 \theta_1 \sin^2 \theta_3
+ 3 \lam_b^2 + \lam_\tau^2 ,\cr
16\pi^2 \ddt \ln \lam_s
& = -c^d_i g_i^2 + \lam_t^2 \cos^2 \theta_1 \sin^2 \theta_3
+ 3 \lam_b^2 + \lam_\tau^2
,\cr
16\pi^2 \ddt \ln \lam_b
& = -c^d_i g_i^2 + \lam_t^2 \cos^2 \theta_3 + 6 \lam_b^2 + \lam_\tau^2 ,\cr
16\pi^2 \ddt \ln \lam_e
& = -c^e_i g_i^2 + 3 \lam_b^2 + \lam_\tau^2 ,\cr
16\pi^2 \ddt \ln \lam_\mu
& = -c^e_i g_i^2 + 3 \lam_b^2 + \lam_\tau^2 ,\cr
16\pi^2 \ddt \ln \lam_\tau
& = -c^e_i g_i^2 + 3 \lam_b^2 + 4 \lam_\tau^2 ,\cr
}
\eqn\eeEigvalRGEone
$$
are obtained by
substituting eq.~\eeCKMParam~into eqs.~\eeYukSqRGE,
again neglecting first and second generation
Yukawa coupling contributions to the running.

Although we assumed
$\lam_u^2 \ll \lam_c^2 \ll \lam_t^2$
and
$\lam_d^2 \ll \lam_s^2 \ll \lam_b^2$
in deriving eqs.~\eeAngleRGE~and \eeEigvalRGEone,
we did {\it not} assume that
the mixing angles $\theta_i$ were small.
Because the third generation mixes with the first two,
third generation quarks induce some running
of the mixing angles $\theta_1$ and $\theta_2$
and the ratios of first and second generation quarks,
$ \lam_u / \lam_c $ and $ \lam_d / \lam_s $.
The amount of running of these quantities,
however,
is typically quite small
because of the smallness of $\theta_3 \sim 0.05 $;
they change by less than 0.1\%
from the GUT scale to the electroweak scale
if $\lam_b$, $\lam_t \lsim 1.5$.
We will therefore be justified
in the following
in neglecting terms proportional
to $\sin^2 \theta_3$ on the r.h.s. of the \RG equations
\eeAngleRGE~and \eeEigvalRGEone.

\chapter{Approximate Solutions to the \RG Equations}

In this section,
we find explicit solutions to
the renormalization group equations for
the fermion masses and CKM matrix elements.
To do so, we need to make several approximations.
In deriving the RG equations \eeAngleRGE~and \eeEigvalRGEone~in
the last section,
we neglected the running
caused by the first and second generations
of Yukawa couplings.
We now make the further assumption
that the mixing angles are small.
In this approximation,
the \RG equations for the CKM matrix elements simplify to
$$
16 \pi^2 \ddt \ln \Vij =
\cases{  - \lam_t^2 - \lam_b^2
        & for $\alpha\beta = ub, cb, td,$ and $ts$, \cr
          ~~0
        & for $\alpha\beta = ud, us, cd, cd,$ and $tb$, \cr}
\eqn\eeCKMRGEthree
$$
and the Yukawa eigenvalues satisfy
$$
\eqalign{
16\pi^2 \ddt \ln \lam_{u,c}
&
= -c^u_i g_i^2 + 3\lam_t^2
,\cr
16\pi^2 \ddt \ln \lam_{t~~}
&
= -c^u_i g_i^2 + 6 \lam_t^2 + \lam_b^2
,\cr
16\pi^2 \ddt \ln \lam_{d,s}
&
= -c^d_i g_i^2 + 3 \lam_b^2 + \lam_\tau^2
,\cr
16\pi^2 \ddt \ln \lam_{b~~}
&
= -c^d_i g_i^2 + \lam_t^2 + 6 \lam_b^2 + \lam_\tau^2
,\cr
16\pi^2 \ddt \ln \lam_{e,\mu}
&
= -c^e_i g_i^2 + 3 \lam_b^2 + \lam_\tau^2
,\cr
16\pi^2 \ddt \ln \lam_{\tau~~}
&
= -c^e_i g_i^2 + 3 \lam_b^2 + 4 \lam_\tau^2
.\cr
}
\eqn\eeEigvalRGEtwo
$$
If we define the scaling factors
$$
A_\alpha (\scale)
= \exp
\left[ \ts{
{1 \over 16\pi^2} \int_{\ln \scale}^{\ln \bscale}
c_i^\alpha g_i^2 (\pscale) \d (\ln \pscale)
          } \right]
\eqn\eeADef
$$
and
$$
B_\alpha (\scale)
= \exp
\left[ \ts{
-~{1 \over 16\pi^2} \int_{\ln \scale}^{\ln \bscale}
\lam_\alpha^2 (\pscale) \d (\ln \pscale)
          } \right],
\eqn\eeBDef
$$
then the solutions to eqs.~\eeCKMRGEthree~and \eeEigvalRGEtwo~are given by
$$
\Vij   (\scale) =
\cases{  \bVij \Bt^{-1} \Bb^{-1}
        & for $\alpha\beta = ub, cb, td,$ and $ts$, \cr
          \bVij
        & for $\alpha\beta = ud, us, cd, cd,$ and $tb$, \cr}
\eqn\eeCKMSoln
$$
and
$$
\eqalign{
&
      \lam_{u} (\scale) = \blam_{u} A_u \Bt^3 ,
\quad\qquad \lam_{d} (\scale) = \blam_{d} A_d \Bb^3 \Btau  ,
\quad\qquad \lam_{e} (\scale) = \blam_{e} A_e       \Bb^3 \Btau  ,
\cr
&
      \lam_{c} (\scale) = \blam_{c} A_u \Bt^3 ,
\quad\qquad \lam_{s} (\scale) = \blam_{s} A_d \Bb^3 \Btau  ,
\quad\qquad \lam_{\mu} (\scale) = \blam_{\mu} A_e       \Bb^3 \Btau  ,
\cr
&
      \lam_{t} (\scale) = \blam_{t} A_u \Bt^6 \Bb ,
\qquad \lam_{b} (\scale) = \blam_{b} A_d  \Bt  \Bb^6 \Btau  ,
\qquad \lam_{\tau} (\scale) = \blam_{\tau} A_e  \Bb^3 \Btau^4 ,
\cr
}
\eqn\eeEigvalSoln
$$
where the overline denotes quantities
evaluated at the GUT scale.
The $A_\alpha$ factors encapsulate the running
induced by the gauge couplings;
the $B_\alpha$ factors that
induced by the Yukawa couplings.

\REF\rrOP{
M.~Olechowski and S.~Pokorski,
\PLB 257 \rm (1991) 388.}
\REF\rrUni{
V.~Barger, M.~S.~Berger, and P.~Ohmann,
Madison preprint MAD/PH/722.}
In the approximation of small mixing angles,
the four off-diagonal CKM matrix elements involving the
third generation all run with the same scale factor
$\Bt^{-1} \Bb^{-1}$,
while the remaining five matrix elements do not run at all,
as has been observed previously [\rrBS, \rrOP, \rrUni].
Also in this approximation,
the ratios of first and second generation Yukawa couplings,
$ \lam_u / \lam_c $,
$ \lam_d / \lam_s $,
and
$ \lam_e / \lam_\mu $,
are invariant under running induced by
third generation Yukawa couplings,
as well as under running induced by gauge couplings [\rrBS, \rrOP, \rrUni].
These results hold to all orders in perturbation theory [\rrUni].

In order to use eqs.~\eeCKMSoln~and \eeEigvalSoln~to scale
fermion mass and mixing angle relations
from the GUT scale
to the supersymmetry breaking scale,
we must know the values of
$A_\alpha (\msusy)$ and $B_\alpha (\msusy)$.
The scaling factors due to the gauge couplings
$A_\alpha (\scale)$ obey
$$
16 \pi^2 \ddt \ln A_\alpha
= -c_i^\alpha g_i^2 ,
\eqn\eeARGE
$$
and are easily calculated using eqs.~\eeGaugeRGE~and \eeGaugeSoln:
$$
A_\alpha (\scale) = \prod_{i=1}^3
\left[ g_i (\scale) \over \bg \right]^{-c_i^\alpha / b_i}
= \prod_{i=1}^3
\left[
1 - {b_i \bg^2 \over 8 \pi^2 } \ln \left( \scale \over \bscale \right)
\right]^{c_i^\alpha / 2b_i}  .
\eqn\eeASoln
$$
These factors are given
at the supersymmetry breaking scale by
$$
A_u (\msusy )= 3.21,
\qquad
A_d (\msusy )= 3.13,
\qquad
A_e (\msusy )= 1.48,
\qquad
{\rm for~} \msusy = 170 \GeV,
\eqn\eeANumer
$$
using eqs.~\eeGaugeRGECoeff~and \eeGaugeGUT.
The scaling factors due to the Yukawa couplings
$B_\alpha (\scale)$ obey
$$
\eqalign{
16 \pi^2 \ddt \ln \Bt
& = \lam_t^2  = \blam_t^2 A_u^2 \Bt^{12} \left[  \Bb^2 \right]
,\cr
16 \pi^2 \ddt \ln \Bb
& = \lam_b^2  = \blam_b^2 A_d^2 \Bb^{12} \left[ \Bt^2 \Btau^2 \right]
,\cr
16 \pi^2 \ddt \ln \Btau
& = \lam_\tau^2  = \blam_\tau^2 A_e^2 \Btau^{12}
\left[ (\Bb / \Btau)^4 \Bb^2 \right]
.\cr
}
\eqn\eeBRGE
$$
The $B_\alpha$ are equal to 1 at the GUT scale,
and decrease monotonically as one lowers the scale.
The equations \eeBRGE~do not have an analytic solution.
We can obtain an approximate solution
by setting the factors in brackets
equal to 1.
The equations then decouple from one another,
and have the solutions
$$
\eqalign{
\Bt (\scale)
& \approx \left[  1 + \blam_t^2 K_u (\scale) \right]^{-1/12} ,
\crr
\Bb (\scale)
& \approx  \left[  1 + \blam_b^2 K_d (\scale) \right]^{-1/12} ,
\qquad\qquad
K_\alpha (\scale) = {3 \over 4\pi^2}
\int_{\ln \scale}^{\ln \bscale}  A_\alpha^2 (\pscale) \d (\ln \pscale).
\crr
\Btau (\scale)
& \approx  \left[  1 + \blam_\tau^2 K_e (\scale) \right]^{-1/12} ,
\cr
}
\eqn\eeBSolnone
$$
We numerically integrate $K_\alpha (\scale)$ to find
$$
K_u (\msusy ) = 8.65,
\qquad
K_d (\msusy )= 8.33,
\qquad
K_e (\msusy )= 3.77,
\qquad
{\rm for~} \msusy = 170 \GeV.
\eqn\eeINumer
$$
The terms in the brackets in eq.~\eeBRGE~are all less than 1,
so omitting them tends to increase the running of the factors $B_\alpha$.
Consequently, the expressions in eq.~\eeBSolnone~are smaller
than the exact values of $B_\alpha$
at $\scale = \msusy$ by about 1 or 2\%.
As we will see in sect.~4,
this approximation tends to exaggerate the effect of the running
induced by the Yukawa couplings.

In the limit $\lam_b$, $\lam_\tau \ll \lam_t$,
the approximate solutions \eeBSolnone~reduce
to the exact result for the running
induced by the top quark alone [\rrDHR, \rrBBHZ]
$$
\Bt (\scale)
= \left[  1 + \blam_t^2 K_u (\scale) \right]^{-1/12} ,
\qquad\qquad
\Bb (\scale)
= \Btau (\scale) = 1.
\eqn\eeBExactSoln
$$

\chapter{Running Relations}

We examine in this section several different
sets of fermion mass and mixing angle relations,
and the effect of renormalization group running on
these relations.
We assume that physics at the GUT scale dictates
certain forms, or textures,
for the matrices of Yukawa couplings.
Different physics leads to different textures.
In this section, we focus on three different textures:
the Georgi-Jarlskog texture,
the Giudice texture,
and the Fritzsch texture.
We will not be concerned so much
with the physics behind these textures,
but simply take them as given,
and examine the relations among fermion masses and mixing angles
to which they give rise.

The relations derived from these various texures
hold at the GUT scale,
and need to be scaled down to low energy to yield predictions
for measured parameters.
We use the spectrum of
the minimal supersymmetric standard model
to run the relations
from the GUT scale
down to the scale at which supersymmetry is broken,
$\msusy=170$ GeV.
Below $\msusy$,
the CKM matrix elements do not evolve much,
\foot{The $t$, $b$, and $\tau$ Yukawa couplings continue
to induce running down to the scale of their masses,
but the amount of running is much less than that
from $\bscale$ to $\msusy$.}
but the Yukawa eigenvalues continue to run due to
QED and QCD effects.
This additional running is incorporated
in the factors $\eta_\alpha$,
defined by
$$
\eta_\alpha = { \lam_\alpha (m_\alpha) \over \lam_\alpha (\msusy)  }.
\eqn\eeEtaDef
$$
In this paper, $m_\alpha$ denotes not the physical mass
but rather the running mass of the fermion,
defined by
$$
m_\alpha = \lam_\alpha (m_\alpha) {v \over \sqrt{2} }
\cases{ \sin \beta & for $\alpha=u$, $c$, and $t$, \cr
        \cos \beta & for $\alpha=d$, $s$, $b$, $e$, $\mu$, and $\tau$. \cr}
\eqn\eeMassDef
$$
The physical (pole) mass of the top quark is then related
to its running mass by
$$
m_t^{\rm phys} =
\left[ 1 + {4 \over 3 \pi} \alpha_3 (m_t) + O (\alpha_3^2) \right] m_t.
\eqn\eeMtopPole
$$
In eqs.~\eeEtaDef~and \eeMassDef,
$\lam_\alpha (m_\alpha)$
should be replaced by
$\lam_\alpha (1 \GeV)$
for the three lightest quarks,
$\alpha = u$, $d$, and $s$.

\REF\rrGL{
J.~Gasser and H.~Leutwyler,
\PRp 87 \rm (1982) 77.
}
When specific numerical values are required
in the following, we will use [\rrGL]
$$
m_\tau = 1.7841^{+0.0027}_{-0.0036} \GeV, \qquad
m_c = 1.27 \pm 0.05 \GeV, \qquad
m_b = 4.25 \pm 0.10 \GeV,
\eqn\eeMassNumer
$$
for the masses, and [\rrBBHZ]
$$
\eta_u = 2.17, \qquad
\eta_s = 2.16, \qquad
\eta_c = 1.89, \qquad
\eta_b = 1.47, \qquad
\eta_\tau = 1.02,
\eqn\eeEtaNumer
$$
for the QCD/QED scaling factors,
corresponding to $\alpha_3 (M_Z) = 0.111$.
We will generally assume that
$m_t$ is close enough to
$\msusy = 170 \GeV$
that running between the two
scales is small,
$$
\eta_t \approx 1.
\eqn\eeEtatNumer
$$
There is considerable uncertainty in the values of the
scaling factors \eeEtaNumer~due
to the uncertainty in $\alpha_3 (M_Z)$ [\rrBBO, \rrADHR].
Since our results are analytic,
it is easy to determine the effects of
choosing other values for $m_\alpha$ and $\eta_\alpha$.

\section{The Georgi-Jarlskog Texture}

The first texture for the Yukawa matrices
we consider is
$$
\MU = \pmatrix{  0 		& C 		& 0 	\cr
                 C 		& 0 		& B 	\cr
  		 0 		& B 		& A 	\cr},
\qquad
\MD = \pmatrix{  0		& F \e^{i\phi}  & 0 	\cr
                 F \e^{-i\phi}  & E		& 0	\cr
  		 0 		& 0 		& D 	\cr},
\qquad
\ME = \pmatrix{  0		& F 		& 0 	\cr
                 F 		& -3E		& 0	\cr
  		 0 		& 0 		& D 	\cr},
\eqn\eeGJTexture
$$
assumed to hold at the grand unification scale.
Georgi and Jarlskog [\rrGJ]
originally posited
this form for the Yukawa matrices
in an SU(5) grand unified theory,
and Harvey, Ramond, and Reiss [\rrHRR]
used it in an SO(10) model.
Recently,
a number of authors [\rrFla, \rrDHR, \rrBBHZ, \rrBBO, \rrADHR]
have re-examined this texture
in a supersymmetric context.
The relations between the $\MD$ and $\ME$ matrix elements
follow if the charged leptons and charge $-\third$ quarks
belong to the same grand unified representation.
Entries of the two matrices that are equal in magnitude
result from Yukawa couplings to a Higgs field
in the 5 of SU(5)
or the 10 of SO(10).
Entries differing by a factor of $-3$
result from Yukawa couplings to a Higgs field
in the 45 of SU(5)
or the 126 of SO(10).
The zero entries of $\MU$, $\MD$, and $\ME$
are due to discrete symmetries [\rrDisc, \rrFri]
at the grand unified scale.

The Georgi-Jarlskog texture  \eeGJTexture~leads to
six relations among fermion masses and mixing angles.
The eigenvalues of the Yukawa matrices \eeGJTexture~obey
the SU(5) relation [\rrSUfive]
$$
{\blam_b \over \blam_\tau} = 1
\eqn\eeGJSUGUT
$$
and the Georgi-Jarlskog relations [\rrGJ]
$$
{\blam_\mu - \blam_e \over \blam_s - \blam_d} = 3 ,
\qquad\qquad
{\blam_e \blam_\mu \over  \blam_d \blam_s} = 1.
\eqn\eeGJRoneGUT
$$
The latter two equations can be combined into
$$
{ \left( \blam_d /  \blam_s \right) \over
  \left[ 1 - \left( \blam_d / \blam_s \right)\right]^2 }
=
{ 9 \left( \blam_e / \blam_\mu \right) \over
  \left[ 1 - \left( \blam_e / \blam_\mu \right) \right]^2 }.
\eqn\eeGJRtwoGUT
$$
The quark Yukawa matrices $\MU$ and $\MD$
are diagonalized by bi-unitary transformations
$  \MFd = \LF^\dag \MF \RF $,
with
$$
\LU = 	    \pmatrix{ 1		& 0 		& 0 		\cr
		      0		& \bcthr	& -\bsthr	\cr
		      0		& \bsthr 	& \bcthr	\cr}
	    \pmatrix{ \bctwo 	& -\bstwo 	& 0 		\cr
		      \bstwo 	& \bctwo 	& 0		\cr
		      0		& 0		& 1		\cr},
\qquad
\LD
	=    \pmatrix{ \phase 	& 0		& 0 		\cr
			0 	& 1		& 0		\cr
		        0	& 0		& 1		\cr}
	    \pmatrix{ \bcone	& -\bsone 	& 0 		\cr
		      \bsone 	& \bcone 	& 0		\cr
		      0		& 0		& 1		\cr},
\eqn\eeGJUnit
$$
and with $\RU$ and $\RD$ equal to $\LU$ and $\LD$
respectively,
modulo diagonal matrices of signs
to make the eigenvalues $\blam_\alpha$ positive.
In the unitary matrices \eeGJUnit,
$\bc_i = \cos \btheta_i$ and $\bs_i = \sin \btheta_i$,
with
$$
\tan^2 \btheta_1
 = {\blam_d \over \blam_s} ,
\qquad\qquad
 \tan^2 \btheta_2
 = {\blam_u \over \blam_c} ,
\qquad\qquad
\tan^2 \btheta_3
 = {\blam_c - \blam_u \over \blam_t} \approx {\blam_c \over \blam_t} .
\eqn\eeGJAngles
$$
The unitary transformations \eeGJUnit~result in a CKM matrix
of exactly the form \eeCKMParam
$$
\bV = \LU^\dag \LD =
\pmatrix{ \bsone \bstwo \bcthr + \bcone \bctwo \phase  &
	  \bcone \bstwo \bcthr - \bsone \bctwo \phase  &
          \bstwo \bsthr \cr
          \bsone \bctwo \bcthr - \bcone \bstwo \phase  &
	  \bcone \bctwo \bcthr + \bsone \bstwo \phase  &
	  \bctwo \bsthr \cr
          -\bsone \bsthr &
	  -\bcone \bsthr &
 	  \bcthr \cr}.
\eqn\eeGJCKMMat
$$
Therefore, in the approximation that the mixing angles are small,
the CKM matrix elements satisfy the
Harvey-Ramond-Reiss (HRR) relation [\rrHRR]
$$
\bVcb \approx \sqrt{\blam_c \over \blam_t} ,
\eqn\eeGJVcbGUT
$$
which depends on the top quark Yukawa coupling,
as well as the relations
$$
\bVus \approx
\left| \sqrt{\blam_d \over \blam_s}
- \e^{-i\phi} \sqrt{\blam_u \over \blam_c} \right|,
\qquad\qquad
{\bVub \over \bVcb} \approx
\sqrt{\blam_u \over \blam_c} ,
\eqn\eeGJCKMGUT
$$
which depend on only
the first and second generation Yukawa couplings.
The six relations \eeGJSUGUT--\eeGJRtwoGUT,
\eeGJVcbGUT, and \eeGJCKMGUT~hold at the GUT scale.

The Yukawa matrices do not retain
the form \eeGJTexture~below the GUT scale
because of \RG running
induced by large Yukawa couplings.
Rather than follow the evolution of the Yukawa matrices,
however, we will determine the effect of the running
on the relations
\eeGJSUGUT--\eeGJRtwoGUT, \eeGJVcbGUT, and \eeGJCKMGUT.
We use eqs.~\eeCKMSoln~and \eeEigvalSoln~to scale
the relations from
the GUT scale $\bscale$ to the supersymmetry breaking scale $\msusy$.
(Whenever the scaling factors $A_\alpha$ and $B_\alpha$
are written without an explicit scale $\scale$ throughout this section,
$\scale = \msusy$ is understood.)
The further running of the Yukawa couplings
from $\msusy$ to the scale of the fermion masses
is included in the factors $\eta_\alpha$
defined in eq.~\eeEtaDef.
Thus, the SU(5) relation \eeGJSUGUT~leads to
$$
{m_b \over m_\tau}
= { A_d \eta_b \over A_e \eta_\tau } { \Bt \Bb^3 \over \Btau^3} ,
\eqn\eeGJSUPhys
$$
and the Georgi-Jarlskog relations \eeGJRoneGUT~and \eeGJRtwoGUT~imply
$$
{m_e m_\mu \over m_d m_s}
= { A_e^2 \eta_e \eta_\mu \over A_d^2 \eta_d \eta_s } ,
\qquad\qquad
{ \left( m_d /  m_s \right) \over
  \left[ 1 - \left( m_d / m_s \right)\right]^2 }
 =
{ 9 \left( \eta_\mu m_e / \eta_e m_\mu \right) \over
  \left[ 1 - \left( \eta_\mu m_e / \eta_e m_\mu \right) \right]^2 },
\eqn\eeGJRPhys
$$
since $\eta_d = \eta_s$.
The HRR relation \eeGJVcbGUT~implies
$$
\Vcb \approx  \sqrt{\eta_t m_c \over \eta_c m_t} \sqrt{\Bt \over \Bb},
\eqn\eeGJVcbPhys
$$
and the CKM relations \eeGJCKMGUT~lead to
$$
\Vus \approx
\left| \sqrt{m_d \over m_s}
- \e^{-i\phi} \sqrt{\eta_c  m_u \over \eta_u m_c } \right|,
\qquad\qquad
{\Vub \over \Vcb} \approx
\sqrt{\eta_c  m_u \over \eta_u m_c } .
\eqn\eeGJCKMPhys
$$
The running induced by the third generation of fermions
is contained in the factors $\Bt$, $\Bb$, and $\Btau$.

Dimopoulos, Hall, and Raby [\rrDHR]
found that four of the six relations implied by the Georgi-Jarlskog texture,
\viz,
the two Georgi-Jarlskog relations \eeGJRoneGUT~and
the two CKM matrix element relations \eeGJCKMGUT,
are unaffected by the running induced by the top quark,
in the limit that the mixing angles are small.
We see from eqs.~\eeGJRPhys~and \eeGJCKMPhys~that,
not surprisingly,
these four relations are also insensitive to the running induced
by the bottom quark and $\tau$ lepton.
Therefore, the predictions of ref.~[\rrDHR] for
$m_s$, $m_d$, $\phi$, and $\left| V_{ub} / V_{cb} \right|$
remain unchanged
even when all three third generation Yukawa couplings contribute
significantly to the running.

The SU(5) and HRR relations,
however,
are modified by the running induced by the third generation of fermions.
The scaling factors $\Bt$, $\Bb$, and $\Btau$
are determined by the \RG equations
\eeBRGE,
which have the approximate solutions
$$
\Bt
\approx  \left[  1 + \blam_t^2 K_u \right]^{-1/12} ,
\qquad
\Bb
\approx  \left[  1 + \blam_b^2 K_d \right]^{-1/12} ,
\qquad
\Btau
\approx  \left[  1 + \blam_\tau^2 K_e \right]^{-1/12} ,
\eqn\eeBSolntwo
$$
where
$K_\alpha  = K_\alpha (\msusy)$
are given in eq.~\eeINumer.
The scaling factors $B_\alpha$
depend on the GUT scale Yukawa couplings
$\blam_t$, $\blam_b$, and $\blam_\tau$.
These couplings are related to
the fermion masses by
$$
\eqalignno{
  m_\tau
& = {v A_e \eta_\tau \over \sqrt{2} }
\;  \blam_\tau  \Bb^3 \Btau^4 (\cos\beta),
& \eqname \eeMtauDef
\cr
m_b
& = {v A_d \eta_b \over \sqrt{2} }
\; \blam_b \Bt \Bb^6 \Btau  (\cos \beta),
& \eqname \eeMbottomDef
\cr
  m_t
& = {v A_u \eta_t \over\sqrt{2}}
\; \blam_t \Bt^6 \Bb  (\sin \beta)
 \approx {v A_u \eta_t \over\sqrt{2K_u}}
  \Bb  \sqrt{ 1 - \Bt^{12} } ~(\sin \beta),
& \eqname \eeMtopDef
\cr
}
$$
using eqs.~\eeEigvalSoln, \eeEtaDef, and \eeMassDef,
where the last equality in eq.~\eeMtopDef~depends on
the approximation~\eeBSolntwo.

The three GUT scale Yukawa couplings
$\blam_t$, $\blam_b$, and $\blam_\tau$
are not independent, however.
First, the Georgi-Jarlskog texture dictates that
$\blam_\tau = \blam_b$.
Second, from the SU(5) relation \eeGJSUPhys, we have
$$
\Bt
= k \left( \Btau \over \Bb \right)^3 ,
\qquad\qquad k
\equiv
{A_e \eta_\tau  m_b \over A_d \eta_b m_\tau } \approx  0.78.
\eqn\eeGJBtop
$$
(The deviation of $k$ from unity shows
that significant running must be induced
by the Yukawa couplings for the SU(5) relation
to be valid.)
Equation \eeGJBtop, together with eq.~\eeBSolntwo,
may be used to determine $\blam_t$ in terms of $\blam_b$.
Hence, $\Bt$, $\Bb$, and $\Btau$,
as well as the fermion masses,
may be written in terms of
a single GUT scale parameter $\blam_b$.
This implies a relation between $\beta$ and $m_t$.
The parameter $\beta$ may be expressed in terms of $\blam_b$ as
$$
\sec \beta
 = {v A_e \eta_\tau \over \sqrt{2} m_\tau} ( \blam_b \Bb^3 \Btau^4 ) ,
\eqn\eeGJBeta
$$
using eq.~\eeMtauDef~and $\blam_\tau = \blam_b$.
The top quark mass is given by
$$
m_t \approx {v A_u \eta_t \over \sqrt{2 K_u} } \Bb
\sqrt{ 1 - k^{12} \left( \Btau / \Bb \right)^{36} }
{}~(\sin \beta),
\eqn\eeGJMtop
$$
using eqs.~\eeMtopDef~and \eeGJBtop.
Plotting  \eeGJBeta~and \eeGJMtop~parametrically
as functions of $\blam_b$,
we obtain the relation between $m_t$ and $\tb$
shown by the solid line in fig.~1.
To see more explicitly the dependence of $m_t$
on $\beta$,
we neglect terms of $O(\blam_b^3)$ on
the r.h.s. of eq.~\eeGJBeta~to obtain
$$
\blam_b \approx {\sec \beta \over 150 } ,
\eqn\eeGJBetaApprox
$$
then expand eq.~\eeGJMtop~in terms of $\blam_b$
to obtain the approximate relation
$$
m_t \approx (185 \GeV) (\sin \beta)
[ 1 - (5 \times 10^{-5}) \sec^2 \beta + \ldots ].
\eqn\eeGJMtopApprox
$$
The Georgi-Jarlskog texture implies an upper bound
on the top quark mass, $m_t \lsim 185 \GeV$,
which is saturated for $\tb \sim 10$.

If the running induced by the $b$ quark and $\tau$ lepton
is neglected,
$m_t$ increases monotonically with $\beta$,
as shown by the dotted line in fig.~1,
and eq.~\eeGJMtopApprox~reduces
to the linear relation
between $ m_t $ and $ \sin \beta $
found in refs. [\rrDHR, \rrBBHZ].
The inclusion of the effects of
all three third generation Yukawa couplings
is responsible for the last factor in eq.~\eeGJMtopApprox,
and causes $m_t$ to decrease for large values of $\tb$.
Thus, a given value of $m_t$ may correspond to two possible
values of $\beta$ [\rrBBO, \rrADHR].

The relation between $m_t$ and $\tb$ obtained by
numerically integrating the \RG equations \eeBRGE~without
approximation is shown by the dashed line in fig.~1,
and is similar to plots shown in
refs.~[\rrFla, \rrKLN, \rrBBO, \rrADHR].
The solid line, based on the approximation~\eeBSolntwo,
has the same qualitative behavior as the dashed line,
but the effects of the Yukawa-induced running
for large $\tb$ are exaggerated,
as anticipated in sect.~3.

The HRR relation \eeGJVcbPhys~determines
the dependence of $\Vcb$ on $\beta$.
Using eq.~\eeGJVcbPhys~together with eq.~\eeGJBtop,
we find
$$
\Vcb
\approx
\sqrt{ A_e \eta_\tau \eta_t m_b    m_c \over
       A_d \eta_b    \eta_c  m_\tau m_t }
\sqrt{\Btau^3 \over \Bb^4}.
\eqn\eeGJVcbPhystwo
$$
(When running induced by the $b$ quark and $\tau$ lepton are neglected,
this reduces to eq.~(22) of ref.~[\rrDHR].)
Using eqs.~\eeGJBeta~and \eeGJVcbPhystwo,
we plot $\Vcb$
as a function of $\tb$ in fig.~2 (solid line).
As before, the dashed line indicates the result
based on numerical integration of the \RG equations
(similar to plots in ref.~[\rrADHR]).
Expanding eq.~\eeGJVcbPhystwo~in terms of $\blam_b$,
we obtain the approximate relation
$$
\Vcb \approx { 0.053 \over \sqrt{ \sin \beta } }
[ 1 + (7 \times 10^{-5} ) \sec^2 \beta + \ldots ].
\eqn\eeGJVcbApprox
$$
The Georgi-Jarlskog texture implies that
$\Vcb$ must be $ \gsim 0.053 $.

In principle, given $\Vcb$,
we could use eq.~\eeGJVcbPhystwo~to determine $\beta$
(up to a two-fold ambiguity)
and therefore $m_t$.
The uncertainty in $\Vcb$,
however, allows us only to set bounds on $\beta$.
For example, requiring $\Vcb < 0.058 $ leads to a lower bound on $\tb$
[\rrDHR, \rrBBHZ]
$$
\Vcb < 0.058
\quad \Rightarrow \quad
\tb \gsim 1.5
\quad \Rightarrow \quad
m_t \gsim 155 \GeV,
\eqn\eeGJBetaLow
$$
as well as an upper bound [\rrBBO, \rrADHR]
$$
\Vcb < 0.058
\quad \Rightarrow \quad
\tb \lsim 40
\quad \Rightarrow \quad
m_t \gsim  170 \GeV.
\eqn\eeGJBetaHigh
$$
It is the running induced by the $b$ quark and $\tau$ lepton
that causes $\Vcb$ to increase for large $\tb$ \eeGJVcbApprox,
and therefore allows us to derive this upper bound on $\tb$;
neglecting this running leads to the relation shown by the dotted
line in fig.~2, which obeys $\Vcb < 0.058$ for all
$\tb \gsim 1.5$.

Finally, we plot $\Vcb$ as a function of $m_t$
in fig.~3.
Here our  analytic approximation (solid line)
almost exactly coincides with the numerical solution (dashed line).
Each value of $\Vcb \gsim 0.053$
corresponds to two different values of $m_t$ [\rrADHR].
The lower branch of the curve is approximately described by
the relation
$$
{ \Vcb \over 0.053 } \approx \sqrt { 185 \GeV \over m_t },
\eqn\eeGJLowerBranch
$$
and holds for $\tb \lsim 8$.
The upper branch of the curve holds for $\tb \gsim 8$;
here the relation between $\Vcb$ and $m_t$ is approximately linear
$$
\Vcb \approx 0.053 + 0.075 \left( 1 - {m_t\over 185 \GeV} \right),
\eqn\eeGJUpperBranch
$$
which we obtain by combining eqs.~\eeGJMtopApprox~and \eeGJVcbApprox.

\section{The Giudice Texture }

Next we consider the Giudice texture  [\rrGiu]
for the Yukawa matrices,
$$
\MU = \pmatrix{  0 		& 0 		& B 	\cr
                 0 		& B 		& 0 	\cr
  		 B 		& 0  		& A 	\cr},
\qquad
\MD = \pmatrix{  0		& F \e^{i\phi}  & 0 	\cr
                 F \e^{-i\phi}  & E		& D	\cr
  		 0 		& D 		& C 	\cr},
\qquad
\ME = \pmatrix{  0		& F 		& 0 	\cr
                 F 		& -3E		& D	\cr
  		 0 		& D 		& C 	\cr},
\eqn\eeGiuTexture
$$
at the GUT scale.
(Giudice additionally imposes the {\it ad hoc} relation $D=2E$,
but we leave these matrix elements unrelated.)
This texture leads to six relations among fermion
mases and mixing angles.
The eigenvalues of the Yukawa matrices
obey not only the Georgi-Jarlskog and SU(5) relations
$$
{\blam_e \over \blam_d} \approx {1 \over 3},
\qquad\qquad
{\blam_\mu \over \blam_s} \approx 3,
\qquad\qquad
{\blam_b \over \blam_\tau} = 1,
\eqn\eeGiuEigvalGUT
$$
but also the geometric mean relation
$$
\blam_u \blam_t = \blam_c^2 .
\eqn\eeGeomMeanGUT
$$
The quark Yukawa matrices are diagonalized by
$$
\LU =       \pmatrix{ \bctwo	& 0 		& -\bstwo 	\cr
		      0		& 1 		& 0 		\cr
		      \bstwo 	& 0 		& \bctwo	\cr},
\qquad
\LD = 	    \pmatrix{ \phase	& 0 		& 0 		\cr
		      0		& \bcthr	& \bsthr	\cr
		      0		& -\bsthr 	& \bcthr	\cr}
	    \pmatrix{ \bcone	& -\bsone 	& 0 		\cr
		      \bsone 	& \bcone 	& 0		\cr
		      0		& 0		& 1		\cr},
\eqn\eeGiuUnit
$$
with
$$
\tan^2 \btheta_1
= {\blam_d \over \blam_s} ,
\qquad\qquad
\tan^2 \btheta_2
= {\blam_u \over \blam_t} ,
\qquad\qquad
\btheta_3
\approx {D \over \blam_b} .
\eqn\eeGiuAngles
$$
The unitary transformations \eeGiuUnit~lead to
a CKM matrix of the form
$$
\bV = \LU^\dag \LD =
\pmatrix{ -\bsone \bstwo \bsthr + \bcone \bctwo \phase  &
	  -\bcone \bstwo \bsthr - \bsone \bctwo \phase  &
          \bstwo \bcthr \cr
          \bsone \bcthr &
	  \bcone \bcthr &
 	  \bsthr \cr
          -\bsone \bctwo \bsthr - \bcone \bstwo \phase  &
	  -\bcone \bctwo \bsthr + \bsone \bstwo \phase  &
	  \bctwo \bcthr \cr
}.
\eqn\eeGiuCKMMat
$$
Thus, for small mixing angles,
the CKM matrix elements obey the relations
$$
\bVus \approx \sqrt{\blam_d \over \blam_s},
\qquad\qquad
\bVub \approx \sqrt{\blam_u \over \blam_t}.
\eqn\eeGiuCKMGUT
$$
The matrix element
$\bVcb = \sin \btheta_3$
remains arbitrary
unless additional constraints [\rrGiu] are imposed on the
parameters of the texture \eeGiuTexture.

The six GUT scale relations
\eeGiuEigvalGUT, \eeGeomMeanGUT~and \eeGiuCKMGUT~are
modified at low energies by \RG running.
Using eqs.~\eeCKMSoln, \eeEigvalSoln, and \eeEtaDef,
we obtain the relations
$$
{ m_e \over m_d}
\approx {1\over 3} {A_e \eta_e \over A_d \eta_d} ,
\qquad\qquad
{m_\mu \over m_s}
\approx 3 { A_e \eta_\mu \over A_d \eta_s}  ,
\qquad\qquad
\Vus \approx \sqrt{m_d \over m_s} ,
\eqn\eeGiuPhysOne
$$
which are not affected
by the running
induced by third generation fermions,
and the relations
$$
\eqalignno{
{m_b \over m_\tau}
& = {A_d \eta_b \over A_e \eta_\tau} {\Bt \Bb^3 \over \Btau^3} ,
& \eqname\eeGiuSUPhys
\cr
m_u
& = { \eta_t \eta_u m_c^2 \over \eta_c^2 m_t } { \Bt^3 \Bb },
& \eqname\eeGeomMeanPhys
\cr
\Vub
&
\approx \sqrt{\eta_t m_u \over \eta_u m_t} \sqrt{ \Bt \over \Bb},
& \eqname\eeGiuVubPhys
\cr
}
$$
which are.
These relations reduce to those given in ref.~[\rrGiu] if
we set $B_b = B_\tau = \eta_\tau = 1$.

The relation between $m_t$ and $\tb$ shown in fig.~1 depends
only  on the SU(5) relation \eeGiuSUPhys,
and therefore holds for the Giudice texture as well
as for the Georgi-Jarlskog texture.
For the same reason,
eqs.~\eeGJBtop--\eeGJMtopApprox~also
continue to hold.

The up quark mass is determined by $\beta$ in the Giudice texture.
The geometric mean relation \eeGeomMeanPhys~together
with eq.~\eeGJBtop~yields
$$
m_u = { \eta_u \eta_t m_c^2 k^3 \over \eta_c^2 m_t }
       \left( \Btau^9 \over  \Bb^8 \right) ,
\eqn\eeGiuMup
$$
which we plot as a function of $\tb$ in fig.~4.
Expanding eq.~\eeGiuMup~in terms of $\blam_b$, we obtain
$$
m_u = { 2.5 \MeV \over \sin \beta }
[ 1 + (1.7 \times 10^{-4}) \sec^2 \beta + \ldots],
\eqn\eeGiuMupApprox
$$
implying a lower bound on the up quark mass, $m_u \gsim 2.5 \MeV$.
We plot the relation between $m_u$ and $m_t$ in fig.~5.
Using eqs.~\eeGJMtopApprox~and \eeGiuMupApprox, we find that
$$
{ m_u\over 2.5 \MeV} \approx  { 185 \GeV \over m_t }
\eqn\eeGiuLowerBranch
$$
for $\tb \lsim 8$,
and
$$
\left( {m_u\over 2.5\MeV} - 1\right)
\approx
3.6 \left( 1 - {m_t\over 185 \GeV} \right),
\eqn\eeGiuUpperBranch
$$
for $\tb \gsim 8$.
The central value
of the up quark mass given in ref.~[\rrGL],
$m_u = 5.1 \pm 1.5 \MeV$,
could correspond either to a small value of $\tb$ ($\sim 0.56$)
with $m_t \sim 90 \GeV $ [\rrGiu],
or to a much larger value of $\tb $ ($\sim 50 $)
with $m_t \sim 130 \GeV $ [\rrBBO].
An up quark in the range $3.6 \MeV < m_u < 6.6 \MeV$
requires either $ 0.4 \lsim \tb \lsim 1.0 $
with $ 70 \GeV \lsim m_t \lsim 130 \GeV $,
or $ 50 \lsim \tb \lsim 60 $
with $ 100 \GeV \lsim m_t \lsim 160 \GeV $.

The Giudice texture also determines $\Vub$ as a function of $\beta$.
{}From eqs.~\eeGJBtop, \eeGeomMeanPhys, and \eeGiuVubPhys,
we have
$$
\Vub \approx { \eta_t m_c k^2 \over \eta_c m_t} {\Btau^6 \over \Bb^6},
\eqn\eeGiuVubPhystwo
$$
which is plotted as a function of $\tb$ in fig.~6.
Expanding eq.~\eeGiuVubPhystwo~in terms of $\blam_b$, we find
$$
\Vub \approx { 0.0022 \over \sin \beta}
[ 1 + (1.5 \times 10^{-4}) \sec^2 \beta + \ldots ],
\eqn\eeGiuVubApprox
$$
implying the lower bound $\Vub \gsim 0.0022$.
The rather weak constraint $\Vub < 0.007$ requires $\tb \gsim 0.3 $.

\section{The Fritzsch Texture }

\REF\rrGN{
F.~Gilman and Y.~Nir,
\ARNPS 40 \rm (1990) 213.
}
Finally, we consider the Fritzsch texture [\rrFri, \rrGN] for the
quark Yukawa matrices,
$$
\MU = \pmatrix{  0 		& C 		& 0 	\cr
                 C 		& 0 		& B 	\cr
  		 0 		& B 		& A 	\cr},
\qquad
\MD = \pmatrix{  0		  & F \e^{i\phi_1}  & 0 	     \cr
                 F \e^{-i\phi_1}  & 0 		    & E \e^{i\phi_2} \cr
  		 0 		  & E \e^{-i\phi_2} & D 	     \cr},
\eqn\eeFriTexture
$$
at the GUT scale.
There is no ansatz for the lepton Yukawa matrix.
The matrices \eeFriTexture~are diagonalized by
$$
\eqalign{
\LU
&         = \pmatrix{ 1		& 0 		& 0 		\cr
		      0		& \bcthrp	& -\bsthrp	\cr
		      0		& \bsthrp 	& \bcthrp	\cr}
	    \pmatrix{ \bctwo 	& -\bstwo 	& 0 		\cr
		      \bstwo 	& \bctwo 	& 0		\cr
		      0		& 0		& 1		\cr},
\crr
\LD
&   =      \pmatrix{ \e^{i\phi_1} & 0 		& 0 		\cr
		      0		   & 1  	& 0		\cr
		      0		   & 0		& \e^{-i\phi_2} \cr}
            \pmatrix{ 1 	& 0 		& 0 		\cr
		      0		& \bcthrpp	& -\bsthrpp	\cr
		      0		& \bsthrpp 	& \bcthrpp	\cr}
	    \pmatrix{ \bcone	& -\bsone 	& 0 		\cr
		      \bsone 	& \bcone 	& 0		\cr
		      0		& 0		& 1		\cr},
\cr
}
\eqn\eeFriUnit
$$
with
$$
\eqalign{
         \tan^2 \btheta_1
&
         = {\blam_d \over \blam_s} ,
         \qquad \qquad
         \tan^2 \btheta^{\prime\prime}_3
       = {\blam_s - \blam_d \over \blam_b}
       \approx {\blam_s \over \blam_b} ,
\crr
         \tan^2 \btheta_2
&    = {\blam_u \over \blam_c} ,
         \qquad\qquad
         \tan^2 \btheta^\prime_3
        = {\blam_c - \blam_u \over \blam_t}
	\approx {\blam_c \over \blam_t} .
\cr
        }
\eqn\eeFriAngles
$$
When the mixing angles are small, the unitary matrices \eeFriUnit~lead to
a CKM matrix
whose elements are given by
$$
\bVus \approx
\left| \sqrt{\blam_d \over \blam_s}
- \e^{-i\phi_1} \sqrt{\blam_u \over \blam_c}
\right|,
\qquad\qquad
{\bVub \over \bVcb} \approx
\sqrt{\blam_u \over \blam_c} ,
\qquad\qquad
\bVcb
\approx \left| \sqrt{\blam_s \over \blam_b }
      - \e^{-i\phi_2} \sqrt{\blam_c \over \blam_t } \right|
\eqn\eeFriCKMGUT
$$
at the GUT scale.
There are no relations among the Yukawa couplings.

We scale down the relations \eeFriCKMGUT~using eqs.~\eeCKMSoln~and
\eeEigvalSoln.
The relations involving only the first and second
generation Yukawa couplings
$$
\Vus \approx
\left| \sqrt{m_d \over m_s}
- \e^{-i\phi_1} \sqrt{\eta_c  m_u \over \eta_u m_c } \right|,
\qquad\qquad
{\Vub \over \Vcb} \approx
\sqrt{\eta_c  m_u \over \eta_u m_c } ,
\eqn\eeFriCKMPhys
$$
are the same as in the Georgi-Jarlskog texture,
and are not affected by the running induced by
large Yukawa couplings.
The relation for $\Vcb$, however, is given by
$$
\Vcb
\approx \left| \sqrt{ \Bb \over \Bt} \sqrt{\eta_b m_s \over \eta_s m_b }
      - \e^{-i\phi_2} \sqrt{ \Bt \over \Bb}
       \sqrt{\eta_t m_c \over \eta_c m_t } \right|
\eqn\eeFriVcbPhys
$$
in the Fritzsch texture.

The relation \eeFriVcbPhys~yields
a connection between
the top quark mass and $\Vcb$.
The small experimental value for $\Vcb$ requires
a large amount of cancellation between the two terms in
eq.~\eeFriVcbPhys,
which seems to imply both $\phi_2 \approx 0$ and
a relatively light top quark [\rrGN].
If we neglect the running induced by Yukawa couplings
and require
$m_s > 120 \MeV$,
$m_b < 4.35 \GeV$ and
$m_c < 1.32 \GeV$ [\rrGL],
we obtain the upper bound for the top quark mass
$$
 \Vcb < 0.058
\quad \Rightarrow \quad
m_t < 110 \GeV,
\qquad\qquad {\rm if~} \Bt=\Bb=1.
\eqn\eeFriTopBound
$$

The inclusion of Yukawa coupling-induced running
implies an even lower upper bound for $m_t$
for small or moderate values of $\tb$,
since typically $B_t < B_b$.
If $\tb$ is very large, however,
then it is possible that $\Bb < \Bt$,
loosening the bound on the top quark,
as Babu and Shafi have pointed out [\rrBS].
They demonstrated this numerically for a specific value of $\tb$.
We will derive a modified bound for $m_t$
using the analytic approximation \eeBSolntwo.
In this approximation,
the top quark mass is given by
$$
m_t
\approx {v A_u \eta_t \over\sqrt{2 K_u} } \Bb
\sqrt{ 1 - \Bt^{12} }  \qquad \qquad {\rm for~} \tb \gg 1.
\eqn\eeMtopDeftwo
$$
The largest amount of cancellation
will occur in eq.~\eeFriVcbPhys~when $\Bb/\Bt$ is minimized.
Holding $m_t$ fixed in eq.~\eeMtopDeftwo,
we find that $\Bb/\Bt$ is minimized when
$ \Bt = \left( 1\over 7 \right)^{1/12} \approx 0.85 $,
and therefore
$$
\left( \Bb \over \Bt \right)_{\rm min} \approx
{ m_t \over 150 \GeV }.
\eqn\eeFriBbBt
$$
Substituting this into eq.~\eeFriVcbPhys,
we obtain
$$
\Vcb < 0.058
\quad \Rightarrow \quad
m_t \lsim 140 \GeV,
\eqn\eeFriTopBoundtwo
$$
a considerably higher upper bound than eq.~\eeFriTopBound.
This high value of the top quark mass
requires a large value of $\tb$.
{}From eq.~\eeMbottomDef, and using the approximation \eeBSolntwo,
we have
$$
\sec \beta
\approx
{v A_d \eta_b \over m_b \sqrt{2 K_d}} \Bt  \Btau \sqrt{ 1 - \Bb^{12} },
\eqn\eeFriSecant
$$
so that
$$
m_t = 140 \GeV
\quad \Rightarrow \quad
\sec \beta \sim 50 \Btau .
\eqn\eeFriTangent
$$
In the Fritzsch texture, $\blam_\tau$ is unrelated to $\blam_b$
so $\Btau$ is undetermined,
but $0.8 \lsim \Btau \le 1$
for $0 \le \blam_\tau < 2$.

\chapter{Conclusions}

Relations among fermion masses and mixing angles
at the scale of grand unification
are modified at lower energies by renormalization group running
induced by gauge and Yukawa couplings.
In supersymmetric theories,
the $b$ quark and $\tau$ lepton Yukawa couplings,
as well as the $t$ quark coupling,
may cause significant running if $\tb$ is large.

In this paper,
we have analyzed the running of fermion masses and mixing angles
caused by the entire third generation of Yukawa couplings.
We made several approximations along the way.
First, we assumed a hierarchy of masses between generations.
In this approximation,
we derived explicit \RG equations for the mixing angles \eeAngleRGE.
Next, we assumed that the mixing angles in the \RG equations
were small;
the error from this approximation was shown to be less than 0.1\%.
Finally, we made the approximation that the
\RG equations \eeBRGE~for the scaling factors $B_\alpha$
decoupled from one another.
This allowed us to obtain the analytic expressions
\eeBSolnone~for these scaling factors,
which differ from the exact values by no more than 1 or 2\%.

We then used the approximate analytic expressions
for the scaling factors
to determine how running induced by the third generation of fermions
affects the predictions arising from the
GUT scale Yukawa matrices.
We summarize here the results
for each of the three textures considered in this paper,
using the input data \eeMassNumer~and \eeEtaNumer.
It is easy to determine the effect
of choosing different values for the input parameters
because our results are analytic.

The Georgi-Jarlskog texture
incorporates the SU(5) relation \eeGJSUPhys,
which implies the relation between $m_t$ and $\tb$
displayed in fig.~1 and given approximately by \eeGJMtopApprox.
The top quark mass is bounded above by $\sim 185 \GeV$,
the bound being saturated for $\tb \sim 10$.
This texture also implies the HRR relation \eeGJVcbPhys;
this leads to the relation between $\Vcb$ and $\tb$
shown in fig.~2 and given approximately by eq.~\eeGJVcbApprox.
Consequently, $\Vcb$ must be greater than $\sim 0.053$.
Requiring $\Vcb \lsim 0.058$ implies that $ 1.5 \lsim \tb \lsim 40$,
and therefore $ m_t \gsim 155 \GeV$.

The Giudice texture also incorporates the SU(5) relation
and therefore the relation between $m_t$ and $\tb$ shown in fig.~1.
This texture also implies the geometric mean relation \eeGeomMeanPhys,
which leads to the relation between $m_u$
displayed in fig.~4 and given approximately by eq.~\eeGiuMupApprox.
The up quark mass is bounded below by $\sim 2.5 \MeV$.
If the up quark lies in the range
$3.6 \MeV < m_u < 6.6 \MeV$ [\rrGL],
then the Giudice texture implies
that either $ 0.4 \lsim \tb \lsim 1.0 $
with $ 70 \GeV \lsim m_t \lsim 130 \GeV $,
or $ 50 \lsim \tb \lsim 60 $
with $ 100 \GeV \lsim m_t \lsim 160 \GeV $.
Finally the Giudice texture implies the relation
between $\Vub$ and $\tb$ shown in fig.~6 and given
approximately by eq.~\eeGiuVubApprox.

The Fritzsch texture relates $\Vcb$ to $m_t$
through eq.~\eeFriVcbPhys.
For small or moderate values of $\tb$,
this relation leads to an upper bound $m_t \lsim 110 \GeV$,
but for large $\tb$ ($\sim 50$),
the top quark could be as heavy as $140 \GeV$.

All of these textures seem able to accommodate
a top quark mass in the range
preferred by the analysis of electroweak
radiative corrections,
at least for some value of $\tb$.
Which one, or whether any of them,
accurately describes reality
remains an open question.
There also remains the more fundamental question:
what underlying mechanism determines the form of the Yukawa matrices
at the GUT scale?

\FIG\fig{
The relation between $m_t$ and $\tb$ implied by the SU(5) relation
\eeGJSUPhys.
Solid line: approximate analytic solution.
Dashed line: numerical solution.
Dotted line:  $b$ and $\tau$ induced running neglected.
}

\FIG\fig{
The relation between $\Vcb$ and $\tb$
implied by the Georgi-Jarlskog texture.
Solid line: approximate analytic solution.
Dashed line: numerical solution.
Dotted line:  $b$ and $\tau$ induced running neglected.
}

\FIG\fig{
The relation between $\Vcb$ and $m_t$
implied by the Georgi-Jarlskog texture.
Solid line: approximate analytic solution.
Dashed line: numerical solution.
}

\FIG\fig{
The relation between $m_u$ and $\tb$
implied by the Giudice texture.
Solid line: approximate analytic solution.
Dashed line: numerical solution.
Dotted line:  $b$ and $\tau$ induced running neglected.
}

\FIG\fig{
The relation between $m_u$ and $m_t$
implied by the Giudice texture.
Solid line: approximate analytic solution.
Dashed line: numerical solution.
}

\FIG\fig{
The relation between $\Vub$ and $\tb$
implied by the Giudice texture.
Solid line: approximate analytic solution.
Dashed line: numerical solution.
Dotted line:  $b$ and $\tau$ induced running neglected.
}

\endpage

\refout
\figout

\end